\documentclass[]{article}
\usepackage[margin=1in]{geometry}
\usepackage[utf8]{inputenc}
\usepackage{paralist}
\usepackage{textcomp}
\usepackage{paralist}
\usepackage{url}

\title{Revisiting {\Large{EZ}}BFT: A Decentralized Byzantine Fault Tolerant Protocol with Speculation}
\author{ \parbox{3.5 in}{\centering Nibesh Shrestha, Mohan Kumar\\
        Department of Computer Science\\
        Rochester Institute of Technology, NY, USA
        {\tt\small \{nxs4564, mjkvcs\}@rit.edu}
        }
}
\date{}
\begin{document}

\maketitle

\begin{abstract}
In this note\footnote{This note pertains to the work presented in \cite{arun2019ezbft}. The authors of \cite{arun2019ezbft} are working on an extended version of the protocol to address the concerns identified in the note.}, we revisit {\footnotesize {EZ}}BFT\cite{arun2019ezbft} and present safety, liveness and execution consistency violations in the protocol. To demonstrate these violations, we present simple scenarios, involving only four replicas, two clients, and one or two owner changes. We also note shortcomings of the presented TLA$^+$ specification used to model check the proposed protocol.

\end{abstract}

\section{Introduction}
\label{sec:Introduction}
{\footnotesize {EZ}}BFT\cite{arun2019ezbft} presents a leaderless and speculative byzantine fault tolerant (BFT) consensus protocol that requires an optimal $3f+1$ replicas to tolerate at most $f$ malicious replicas. To be leaderless, {\footnotesize {EZ}}BFT adopts protocol design specification from Egalitarian Paxos (EPaxos) \cite{moraru2013there} and assigns a two-dimensional instance space with each replica proposing in their own instance space. Replicas execute the consensus protocol to reach agreement on a proposed command (say, $\alpha$) and its ordering attributes -- non-commutative commands with respect to the proposed command $\alpha$. 

Replicas speculatively (and \textit{optimistically}) execute a proposed command with respect to ordering attributes before reaching agreement, and reply directly to the client as in Zyzzyva\cite{kotla2007zyzzyva}. The clients can also be byzantine faulty. {\footnotesize {EZ}}BFT completes in two communication steps when all $3f+1$ replicas respond identically and there are no failures. The protocol requires additional two communication steps when some replicas fail or the ordering attributes for the proposed command aren't final. In such situations, the client finalizes the ordering attributes and commits the finalized command in two communication steps. This corresponds to a possibly byzantine faulty client committing a command in two communication steps. However, a two-step consensus protocol requires at least $5f+1$ replicas \cite{martin2006fast}. As a result, {\footnotesize {EZ}}BFT violates safety property of a consensus protocol. {\footnotesize {EZ}}BFT also violates liveness and execution consistency; a requirement for total ordering of non-commutative commands.

In this note, we present three simple execution scenarios that exhibit safety, execution consistency, and liveness violations. To demonstrate the scenarios, we require a simple setup with only 4 replicas tolerating $f=1$ byzantine failures, two clients, one of which is byzantine faulty, and one owner change during protocol execution. Owner change is a process that elects a new leader to replace a faulty leader. Although, {\footnotesize {EZ}}BFT has been formally verified in TLA$^+$\cite{lamport2002specifying}, the presented specification doesn't fully specify the proposed protocol. We discuss shortcomings of the specification in Section \ref{sec:model_checking}. 

Shrestha et al.\cite{shrestha2019revisiting} have shown a similar safety violation in hBFT\cite{duan2015hbft}, a consensus protocol that promises two-step executions with only $3f+1$ replicas. 

We have communicated with the authors of the {\footnotesize {EZ}}BFT paper regarding the protocol violations. They have notified us that they are working on an extended version of the protocol to address the presented issues.
\section{Preliminaries}
\label{sec:preliminaries}
{\footnotesize {EZ}}BFT requires a total of $N = 3f+1$ replicas with each replica given a unique id $\in \{1...3f+1\}$. The protocol considers a standard communication channel that is authenticated, reliable and asynchronous; messages sent between replicas may take a long time before being delivered, but are never lost. Replicas exchange messages on the communication channel to reach consensus on a common value. The entire state space for each replica is represented as a two-dimensional instance space. Each replica $R$ is assigned its own instance space that consists of unbounded sequence of numbered instances such as $R.0$, $R.1$, $R.3$,... where only replica $R$ is allowed to propose. At most one command may be chosen at any instance.

{\footnotesize {EZ}}BFT deals with partial ordering of proposed commands. In partially ordered commands, non-commutative commands are totally ordered while commutative commands can be in any order. Non-commutative commands are called interfering commands. To ensure total ordering of interfering commands, the protocol attempts to reach agreement on not only the proposed command, but also its ordering attributes. The ordering attributes include a list of interfering commands (also called command dependencies) and a sequence number. The ordering attributes are used to totally order interfering commands before execution. We represent the command and its ordering attributes as $\langle \alpha, deps_\alpha, seq_\alpha\rangle$ where $\alpha$ represents the proposed command, $deps_\alpha$ represents the set of interfering commands that have been proposed and $seq_\alpha$ represents the sequence number assigned to the command $\alpha$. The sequence number $n$ is computed to be one higher than the maximum of sequence numbers of the interfering commands.

A consensus protocol needs to satisfy following safety and liveness requirements:
\begin{itemize}
    \item \textbf{Agreement.} All correct replicas commit a common command and its ordering attributes for the same instance. This property is also called safety.
    \item \textbf{Validity.} A committed command must have been proposed by a client.
    \item \textbf{Execution Consistency.} If two interfering commands $\gamma$ and $\alpha$ are committed, they will be executed in the same order by all correct replicas.
    \item \textbf{Liveness.} A value proposed by a correct client must eventually be committed given the communication channel is eventually partially synchronous.
\end{itemize}

\subsubsection*{Owner Number}
{\footnotesize {EZ}}BFT executes through a series of owner numbers. The protocol moves to a higher owner number when the protocol fails to make progress at current owner number. A correct replica participates in a single owner number at a time moving from a lower numbered owner number to a higher numbered owner number only. The default owner number for each instance is the replica id of instance owner. A new instance leader is elected when the owner number increments. The owner number uniquely identifies the current leader of the instance and is computed as $O \textit{ mod N}$, where $O$ is the owner number.
\section{Skeletal Overview of {\large {EZ}}BFT}
\label{sec:overview}

{\footnotesize {EZ}}BFT revolves around proposing a command, collecting its interfering commands, computing its ordering attributes and committing the command and its ordering attributes. In {\footnotesize {EZ}}BFT, a command may get committed in two ways:
\begin{itemize}
    \item \textit{Fast} path in which a proposed command and its ordering attributes are committed in two communication steps when the proposed command and its ordering attributes are final and there are no byzantine failures with all $3f+1$ replicas responding in a timely manner.
    \item \textit{Slow} path that requires two additional communication steps in which either the proposed ordering attributes aren't final or less than $3f+1$ replicas respond in a timely manner.
\end{itemize}

\subsection*{The Fast Path Protocol}
The fast path protocol is initiated when a client $c$ sends a command $\alpha$ to a replica $L$. Replica $L$ becomes leader for the client command $\alpha$ and assigns next available instance $L.i$ in its instance space. Replica $L$ enlists the commands that interferes with command $\alpha$ in its entire local instance space. The interfering commands aren't necessarily committed. Replica $L$ also computes a sequence number that is one higher than the maximum of sequence numbers assigned to the interfering commands. Assume, the command and its ordering attributes be $\langle \alpha, deps_\alpha, seq_\alpha\rangle$. 

Replica $L$ sends SPEC-ORDER messages to all replicas to propose tuple $\langle \alpha, deps_\alpha, seq_\alpha\rangle$ at instance $L.i$. A replica (say, $R$) accepts the SPEC-ORDER message if no command has been proposed at instance $L.i$. Replica $R$ updates command dependencies if it has seen any new interfering commands not included in $deps_\alpha$, and updates $seq_\alpha$ accordingly. Replica $R$ speculatively executes command $\alpha$  before the command and its ordering attributes get committed. Command $\alpha$ is executed based on its ordering attributes. See Section IV.E in \cite{arun2019ezbft} for more details on the ordering algorithm. Replica $R$ sends speculative response SPEC-REPLY to the client $c$.

Client $c$ waits for responses from replicas for a certain time duration. If all $3f+1$ replicas sends SPEC-REPLY responses for the same tuple $\langle \alpha, deps_\alpha, seq_\alpha\rangle$ in a timely manner, client $c$ considers the command $\alpha$ to be complete. It also sends COMMIT-FAST messages to all replicas to signal the completion of the command $\alpha$. The COMMIT-FAST message is accompanied by a \textit{commit} certificate that consists of identical $3f+1$ SPEC-REPLY responses. Replicas also commit command $\alpha$ when they receive COMMIT-FAST message with a valid \textit{commit} certificate.

\subsection*{The Slow Path Protocol}
The slow path protocol is triggered when client $c$ times out before receiving all $3f+1$ SPEC-REPLY responses with at least $2f+1$ SPEC-REPLY response or when at least one SPEC-REPLY response contains a tuple with updated ordering attributes. In this case, the client computes the union of all command dependencies to find new $deps_\alpha$, and updates $seq_\alpha$ accordingly. The finalized tuple $\langle \alpha, deps_\alpha, seq_\alpha\rangle$ is sent to all replicas in a COMMIT message. The COMMIT message is accompanied by a \textit{commit} certificate that consists of $2f+1$ SPEC-REPLY responses that vouch for the updated ordering attributes.

A replica accepts the finalized command and its ordering attributes in the COMMIT message if the \textit{commit} certificate vouches for the updated command and its attributes. If the ordering attributes have changed since the last time the replica executed the command $\alpha$, the replica re-executes the command based on new ordering attributes and responds to the client with updated results in a COMMIT-REPLY message. 

The client waits for $2f+1$ identical COMMIT-REPLY messages to consider the command $\alpha$ complete. Here, the client doesn't send any signals to notify the replicas that command $\alpha$ has been completed.

\subsection*{The Owner Change Protocol}
{\footnotesize {EZ}}BFT initiates with a default owner number. The protocol transitions to a new owner number with a new instance leader when the protocol fails to make progress at current owner number. To make the transition, the owner change protocol is executed that selects a safe tuple to propose in a higher owner number. The owner change protocol collects $N-f$ OWNER-CHANGE messages which includes the SPEC-REPLY responses sent by the replicas in previous owner number along with any \textit{commit} certificates a replica has seen.

We present the rule for the owner change protocol in {\footnotesize {EZ}}BFT to select a safe tuple. A tuple P$_i$ is considered safe in one of the following conditions:
\begin{itemize}[]
    \item \textbf{Condition 1:} $P_i$ has a COMMIT message with the highest owner number to prove its entries (command and its ordering attributes).
    \item \textbf{Condition 2:} $P_i$ has at least $f +1$ SPEC-REPLY messages with the highest owner number to prove its entries.
\end{itemize}

If there exists a tuple P$_j$ that extends a P$_i$ satisfying any of the above conditions, then P$_j$ is a valid extension of P$_i$ if one of the following conditions hold:
\begin{enumerate}[1)]
    \item $P_j$ satisfies Condition 1, and for every command $\gamma \in P_j \setminus P_i$, $\gamma$ has at least $f + 1$ SPEC-REPLY messages with the same highest order number as P$_i$.
    \item $P_j$ satisfies Condition 2, and for every command  $\gamma \in P_j \setminus P_i$, $\gamma$ has a COMMIT message with the same highest order number as P$_i$.
\end{enumerate}

The new leader selects a safe tuple based on the above rule and sends the tuple to all replicas in a NEW-OWNER message at a higher owner number. The NEW-OWNER message is accompanied by $N-f$ OWNER-CHANGE messages collected during the owner change protocol. Replicas accept the presented safe tuple if it is vouched by a set of $N-f$ OWNER-CHANGE messages in the NEW-OWNER message. Replicas consider the accepted tuple committed.
\section{Protocol Violations}
\label{sec:violation}
Consider four replicas $R$, $L$, $Q$ and $T$ among which replica $T$ is byzantine. Also, consider two clients $c_1$ and $c_2$, of which client $c_1$ is faulty.

\subsection{Breaking Safety}
In this scenario, we show safety violation  in {\footnotesize {EZ}}BFT with the faulty client $c_1$.

\begin{itemize}[]
    \item \textbf{Client $c_1$ proposing command $\alpha$ and client $c_2$ proposing command $\beta$. Commands $\alpha$ and $\beta$ are interfering in nature.}
    
\begin{enumerate}
    \item Client $c_1$ sends command $\alpha$ to replica $R$. Replica $R$ computes ordering attributes for $\alpha$. Assume replica $R$ has not seen any commands that interfere with $\alpha$. Hence, replica $R$ computes the tuple to be $\langle \alpha, \{\}, 1 \rangle$ and assigns an instance $R.0$.
    
    \item Around the same time, client $c_2$ sends command $\beta$ to replica $T$ which is byzantine.
    Replica $T$ computes tuple $\langle \beta, \{\}, 1 \rangle$ for command $\beta$ and assigns an instance $T.0$.
    
    \item Replica $R$ sends SPEC-ORDER to propose $\langle \alpha, \{\}, 1 \rangle$ to all replicas at instance $R.0$.
    
    \item Assume replicas $L$ and $Q$ have not seen any commands that interfere with $\alpha$; hence, they do not update the proposed tuple. Both replicas speculatively execute $\langle \alpha, \{\}, 1 \rangle$ and send SPEC-REPLY to client $c_1$ for tuple $\langle \alpha, \{\}, 1 \rangle$.
    
    \item Replica $R$ also speculatively executes $\langle \alpha, \{\}, 1 \rangle$ and responds to client $c_1$ with SPEC-REPLY for tuple $\langle \alpha, \{\}, 1 \rangle$.
    
    \item Replica $T$, which is byzantine, has seen command $\beta$ that interferes with $\alpha$. Replica $T$, being byzantine, computes two tuples $\langle \alpha, \{\}, 1 \rangle$ and $\langle \alpha, \{\beta\}, 2 \rangle$ and hence sends two SPEC-REPLY responses to the client $c_1$
    
    \item Client $c_1$ receives responses from all four replicas in a timely manner. Client $c_1$, being faulty, forms two \textit{commit} certificates.
    
    \begin{enumerate}[i.]
        \item \textit{Commit} certificate -- \textit{CC-Fast}, consists of $3f+1$ (i.e., 4) identical SPEC-REPLY responses for $\langle \alpha, \{\}, 1 \rangle$.  \textit{CC-Fast} vouches for tuple $\langle \alpha, \{\}, 1 \rangle$. 
        \item \textit{Commit} certificate -- \textit{CC-Slow}, comprises $f+1$ SPEC-REPLY responses for $\langle \alpha, \{\}, 1 \rangle$ and $f$ SPEC-REPLY responses for $\langle \alpha, \{\beta\}, 2 \rangle$ (from byzantine replica $T$).  \textit{CC-Slow} vouches for tuple $\langle \alpha, \{\beta\}, 2 \rangle$.
    \end{enumerate}
    
    \item Client $c_1$ sends \textit{CC-Fast} to replica $R$ and \textit{CC-Slow} to replica $Q$.
    
    \item Replica $R$ receives \textit{CC-Fast} that consists of all $3f+1$ identical SPEC-REPLY responses. As per {\scriptsize {EZ}}BFT protocol, replica $R$ considers tuple $\langle \alpha, \{\}, 1 \rangle$ final for instance $R.0$ and commits it.
    
    Here, \textbf{Replica $R$ commits on the tuple $\langle \alpha, \{\}, 1 \rangle$.}
    
    \item Replica $Q$ receives \textit{CC-Slow} that vouches for $\langle \alpha, \{\beta\}, 2 \rangle$. Replica $Q$ re-executes the finalized tuple $\langle \alpha, \{\beta\}, 2 \rangle$ and sends COMMIT-REPLY for the finalized tuple to client $c_1$.
    
\end{enumerate}

At this stage, all further messages are delayed and owner change protocol is triggered.  During owner change protocol, replicas send their SPEC-REPLY responses along with any \textit{commit} certificates they have. Assume the next owner for instance $R.0$ is replica $L$.

\begin{enumerate}
    \item Replica $L$ collects  $2f+1$ OWNER-CHANGE messages, all of which are at the highest owner number.
    \begin{itemize}
        \item Replica $L$ sends its accepted tuple $\langle \alpha, \{\}, 1 \rangle$.
        \item Replica $Q$ sends its finalized tuple $\langle \alpha, \{\beta\}, 2 \rangle$ along with the \textit{commit} certificate, \textit{CC-Slow}, for tuple $\langle \alpha, \{\beta\}, 2 \rangle$.
        \item Replica $T$, which is byzantine, sends tuple $\langle \alpha, \{\beta\}, 2 \rangle$.
    \end{itemize}
    Response from replica $R$ gets delayed.
    
    \item As per the protocol specification during owner change (\textbf{Condition 1}), replica $L$ selects tuple $\langle \alpha, \{\beta\}, 2 \rangle$ as a safe tuple as it is vouched by a \textit{commit} certificate, \textit{CC-Slow}.
    
    \item Replica $L$ proposes tuple $\langle \alpha, \{\beta\}, 2 \rangle$ at instance $R.0$ along with a proof that includes \textit{commit} certificate, \textit{CC-Slow} in NEW-OWNER message.
    
    \item Replicas $L$, $Q$ and $T$ receive NEW-OWNER messages that proposes tuple $\langle \alpha, \{\beta\}, 2 \rangle$ with a valid \textit{commit} certificate, \textit{CC-Slow}. Because of a valid proof, they accept tuple $\langle \alpha, \{\beta\}, 2 \rangle$. Replica $L$ re-executes the finalized tuple $\langle \alpha, \{\beta\}, 2 \rangle$ and responds to the client $c_1$.
    
    Here, \textbf{Replicas $L$ and $Q$ commit on the tuple $\langle \alpha, \{\beta\}, 2 \rangle$.}
\end{enumerate}
\end{itemize}

\subsection{Breaking Execution Consistency}
\label{sec:violation_execution_consistency}
In this scenario, we show violation of execution consistency in {\footnotesize {EZ}}BFT. To show the violation, we do not require any byzantine behavior from replicas and clients.

\begin{itemize}[]
    \item \textbf{Client $c_1$ proposing command $\alpha$ and client $c_2$ proposing command $\beta$. Commands $\alpha$ and $\beta$ are interfering in nature.}

\begin{enumerate}
    \item Client $c_1$ sends command $\alpha$ to replica $R$. Replica $R$ computes ordering attributes for $\alpha$. Assume replica $R$ has not seen any commands that interfere with $\alpha$. Hence, replica $R$ computes the tuple to be $\langle \alpha, \{\}, 1 \rangle$ and assigns an instance $R.0$.
    
    \item Similarly, client $c_2$ sends command $\beta$ to replica $Q$. Replica $Q$ computes ordering attributes for $\beta$. Assume replica $Q$ has not seen any commands that interfere with $\beta$. Hence, replica $Q$ computes the tuple to be $\langle \beta, \{\}, 1 \rangle$ and assigns an instance $Q.0$.
    
    \item Replica $R$ sends SPEC-ORDER to all replicas to propose $\langle \alpha, \{\}, 1 \rangle$ at instance $R.0$. Replica $R$ also speculatively executes $\langle \alpha, \{\}, 1 \rangle$ and sends SPEC-REPLY to client $c_1$.
    
    \item Similarly, replica $Q$ sends SPEC-ORDER to all replicas to propose $\langle \beta, \{\}, 1 \rangle$ at instance $Q.0$. Replica $Q$ also speculatively executes $\langle \beta, \{\}, 1 \rangle$ and responds to client $c_2$ with a SPEC-REPLY for $\langle \beta, \{\}, 1 \rangle$.
    
    \item Assume replica $L$ has not seen any commands that interfere with $\alpha$; hence, it does  not update any proposed tuple. Replica $L$ speculatively executes the tuple $\langle \alpha, \{\}, 1 \rangle$ and responds client $c_1$ with SPEC-REPLY for $\langle \alpha, \{\}, 1 \rangle$.
    
    \item Replica $Q$ receives the proposed tuple $\langle \alpha, \{\}, 1 \rangle$ at instance $R.0$. Replica $Q$ has seen command $\beta$ that interfere with command $\alpha$ at instance $Q.0$. It finalizes the tuple to be $\langle \alpha, \{\beta\}, 2 \rangle$. Replica $Q$ speculatively executes the finalized tuple and responds to client $c_1$ with SPEC-REPLY for tuple $\langle \alpha, \{\beta\}, 2 \rangle$.
    
    \item Assume replica $T$ has not seen any commands that interferes with $\beta$. hence, it does  not update any the proposed tuple. Replica $T$ speculatively executes the tuple $\langle \beta, \{\}, 1 \rangle$ and responds to client $c_2$ with SPEC-REPLY for $\langle \beta, \{\}, 1 \rangle$.
    
    \item Replica $R$ receives the proposed tuple $\langle \beta, \{\}, 1 \rangle$ at instance $Q.0$. Replica $R$ has seen command $\alpha$ that interferes with command $\beta$ at instance $R.0$. It finalizes the tuple to be $\langle \beta, \{\alpha\}, 2 \rangle$. Replica $R$ speculatively executes the finalized tuple and responds client $c_2$ with SPEC-REPLY for tuple $\langle \beta, \{\alpha\}, 2 \rangle$.
\end{enumerate}

At this stage, all further messages are delayed and owner change protocol is triggered for both instances $R.0$ and $Q.0$. Assume the next owner for instance $R.0$ is replica $L$ and the next owner for instance $Q.0$ is replica $T$.

\textbf{Recovery for instance $R.0$}
\begin{enumerate}
    \item Replica $L$ collects $2f+1$ OWNER-CHANGE messages, all of which are at the highest owner number.
    \begin{itemize}
        \item Replica $L$ (itself) sends its accepted tuple $\langle \alpha, \{\}, 1 \rangle$.
        \item Replica $R$ sends its accepted tuple $\langle \alpha, \{\}, 1 \rangle$.
        \item Replica $Q$ sends its accepted tuple $\langle \alpha, \{\beta\}, 2 \rangle$.
    \end{itemize}
    Here, there are no any \textit{commit} certificates.
    
    \item As per protocol specification during owner change (\textbf{Condition 2}), replica $L$ selects tuple $\langle \alpha, \{\}, 1 \rangle$ as a safe tuple as there exist $f+1$ SPEC-REPLY messages. Although, tuple $\langle \alpha, \{\beta\}, 2 \rangle$ extends tuple $\langle \alpha, \{\}, 1 \rangle$, there exists no COMMIT message for command $\beta$. Hence,  {\scriptsize {EZ}}BFT computes $\langle \alpha, \{\}, 1 \rangle$ to be the safe tuple.
    
    \item New instance leader $L$ proposes tuple $\langle \alpha, \{\}, 1 \rangle$ at instance $R.0$ along with a proof that shows the proposed tuple is safe in NEW-OWNER message.
    
    \item Replicas $L$, $Q$ and $T$ accept the proposed tuple $\langle \alpha, \{\}, 1 \rangle$ accompanied by a valid proof, speculatively execute the finalized tuple and commit on it.
    
    Here, \textbf{tuple $\langle \alpha, \{\}, 1 \rangle$ is committed at instance $R.0$}.
\end{enumerate}

\textbf{Recovery for instance $Q.0$}
\begin{enumerate}
    \item Replica $T$ collects $2f+1$ owner change messages, all of which are at the highest owner number.
    \begin{itemize}
        \item Replica $T$ (itself) sends its accepted tuple $\langle \beta, \{\}, 1 \rangle$.
        \item Replica $Q$ sends its accepted tuple $\langle \beta, \{\}, 1 \rangle$ for instance $Q.0$.
        \item Replica $R$ sends its accepted tuple $\langle \beta, \{\alpha\}, 2 \rangle$.
    \end{itemize}
    
    \item With similar explanation as above, replica $T$ computes tuple $\langle \beta, \{\}, 1 \rangle$ as safe tuple.
    
    \item New instance leader $T$ proposes tuple $\langle \beta, \{\}, 1 \rangle$ at instance $Q.0$ in NEW-OWNER message.
    
    \item Replicas $R$, $Q$ and $T$ accept the proposed tuple $\langle \beta, \{\}, 1 \rangle$ accompanied by a valid proof, speculatively execute the finalized tuple and commit on it.
    
    Here, \textbf{tuple $\langle \beta, \{\}, 1 \rangle$ is committed at instance $Q.0$}.
\end{enumerate}
Here, interfering commands $\alpha$ and $\beta$ get committed without being in either command's dependency list. Hence, interfering commands $\alpha$ and $\beta$ may be executed in any order violating execution consistency.
\end{itemize}

\subsection{Breaking Liveness}
In this scenario, we show liveness violation. To show the violation, we require a single faulty client $c_1$.

\begin{itemize}[]
    \item \textbf{Client $c_1$ proposing command $\alpha$.}
    
\begin{enumerate}
    \item Client $c_1$ sends command $\alpha$ to replica $R$. Replica $R$ computes ordering attributes for $\alpha$. Replica $R$ computes the tuple to be $\langle \alpha, \{\}, 1 \rangle$ and assigns an instance $R.0$.
    
    \item Replica $R$ sends SPEC-ORDER to all replicas to propose $\langle \alpha, \{\}, 1 \rangle$ at instance $R.0$. Replica $R$ also speculatively executes $\langle \alpha, \{\}, 1 \rangle$ and responds to client $c_1$ with SPEC-REPLY for tuple $\langle \alpha, \{\}, 1 \rangle$.
    
    \item Assume replica $L$ and $Q$ have not seen any commands that interfere with $\alpha$; hence, they do not update the proposed tuple. Both replicas speculatively execute $\langle \alpha, \{\}, 1 \rangle$ and send SPEC-REPLY to client $c_1$ for tuple $\langle \alpha, \{\}, 1 \rangle$.
    
    \item Assume replica $T$ has seen a command $\beta$ (from some client $c_2$) that interfere with command $\alpha$. Replica $T$ updates the tuple to be  $\langle \alpha, \{\beta\}, 2 \rangle$, speculatively executes it and sends SPEC-REPLY message to client $c_1$ for  $\langle \alpha, \{\beta\}, 2 \rangle$.
    
    \item Client $c_1$, which is faulty, forms two \textit{commit} certificates each with $2f+1$ SPEC-REPLY messages.
    \begin{itemize}[i.]
        \item A \textit{commit} certificate \textit{CC1} that contains $2f+1$ identical SPEC-REPLY messages for tuple $\langle \alpha, \{\}, 1 \rangle$.
        \item A \textit{commit} certificate \textit{CC2} that contains $f+1$ SPEC-REPLY messages for tuple $\langle \alpha, \{\}, 1 \rangle$ and $f$ SPEC-REPLY messages for $\langle \alpha, \{\beta\}, 2 \rangle$. \textit{CC2} vouches for $\langle \alpha, \{\beta\}, 2 \rangle$.
    \end{itemize}
    
    \item Client $c_1$ sends \textit{commit} certificate \textit{CC1} to replica $R$ and \textit{commit} certificate \textit{CC2} to replica $L$.
\end{enumerate}

At this stage, all further messages are delayed and owner change protocol is triggered for instance $R.0$. Assume the next owner for instance $R.0$ is replica $L$.

\begin{enumerate}
    \item Replica $L$ collects  $2f+1$ owner change messages, all of which are at the highest owner number.
    \begin{itemize}
        \item Replica $R$ sends its accepted tuple $\langle \alpha, \{\}, 1 \rangle$ along with \textit{CC1} that vouches for tuple $\langle \alpha, \{\}, 1 \rangle$.
        \item Replica $L$ sends its accepted tuple $\langle \alpha, \{\beta\}, 2 \rangle$ along with \textit{CC2} that vouches for tuple $\langle \alpha, \{\beta\}, 2 \rangle$.
        \item Replica $Q$ sends its accepted tuple $\langle \alpha, \{\}, 1 \rangle$.
    \end{itemize}
\end{enumerate}

At this stage, there exist two \textit{commit} certificates, \textit{CC1} and \textit{CC2}, that vouch for different tuples. Both the tuples are at the highest owner number. The owner change protocol in {\footnotesize {EZ}}BFT doesn't specify any rules to handle such scenarios. Hence, the protocol gets stuck violating the liveness property.
\end{itemize}
\section{TLA$^+$ Formal Verification}
\label{sec:model_checking}

{\footnotesize {EZ}}BFT provides TLA$^+$ \cite{lamport2002specifying} specification of the protocol in the technical report \cite{arun2019ezbfttech} and has been model checked using TLC model checker\cite{tlcModelChecker} to verify the correctness of the protocol. However, the specification appears to have following limitations.

\begin{itemize}
    \item {\footnotesize {EZ}}BFT doesn't fully specify how byzantine replicas can behave. In the specification, the byzantine behavior of faulty replicas is restricted to simply replying with empty dependencies and a sequence number of 1. The byzantine leaders also do not propose inconsistent commands which corresponds to actions similar to correct leaders. In reality, byzantine leaders can propose inconsistent commands resulting in inconsistent command dependencies and byzantine replicas may append partial or no dependencies at all.
    
    \item The paper mentions the protocol can handle unlimited number of faulty clients. However, the TLA$^+$ specification doesn't specify any faulty client behavior. The clients always behave correctly in the specification. A faulty client may form different \textit{commit} certificates and send different certificates to different replicas. The scenarios presented above use such faulty client behavior to show safety and liveness violations.
    
    \item The specification doesn't check the condition that if two interfering commands $\alpha$ and $\gamma$ are committed, then either $\alpha$ has $\gamma$ in $\alpha$'s dependency list or $\gamma$ has $\alpha$ in $\gamma$'s dependency list or both will have each other in their dependency list. This condition is required to ensure execution consistency property. The scenario presented in Section \ref{sec:violation_execution_consistency} shows execution consistency violation without any byzantine behavior.
\end{itemize}

\newpage
\bibliographystyle{abbrv}
\bibliography{main}
\end{document}